\documentclass[useAMS,usenatbib]{mn2e}
\usepackage{graphics}

\def\oversim#1#2{\lower0.5pt\vbox{\baselineskip0pt \lineskip-0.5pt
     \ialign{$\mathsurround0pt #1\hfil##\hfil$\crcr#2\crcr\sim\crcr}}}
\begin{document}

\title[Spectra of AGB stars in Fornax dSph Galaxy]
{Spitzer Space Telescope spectral observations of AGB stars in the Fornax dwarf spheroidal galaxy}
 %  \subtitle{Mass-loss rate at 1/10th of the solar metallicity}
\author[M. Matsuura et al.]
{M.~Matsuura$^{1,2}$, 
A.A.~Zijlstra$^{3}$,
J.~Bernard-Salas$^{4}$, 
J.W.~Menzies$^{5}$, 
\newauthor
G.C.~Sloan$^{4}$,
P.A.~Whitelock$^{5,6,7}$,
P.R.~Wood$^{8}$,  
M.-R.L.~Cioni$^{9}$, 
M.W.~Feast$^{6}$, 
\newauthor
E. Lagadec$^{3}$,
J.Th.~van~Loon$^{10}$, 
M.A.T.~Groenewegen$^{11}$,
G.J.~Harris$^{2}$ \\
%
%$^{2}$ APS Division, Department of Pure and Applied Physics, 
%        Queen's University Belfast, University Road, BT7 1NN, United Kingdom \\
$^{1}$ National Astronomical Observatory of Japan, Osawa 2-21-1, 
       Mitaka, Tokyo 181-8588, Japan \\
$^{2}$ Department of Physics and Astronomy, University College London, 
	Gower Street, London WC1E 6BT, United Kingdom \\
$^{3}$ School of Physics and Astronomy, University of Manchester, 
        Sackville Street, P.O. Box 88, Manchester M60 1QD, United Kingdom \\
$^{4}$ Astronomy Department, Cornell University, 610 Space Sciences Building, 
        Ithaca, NY 14853-6801, USA \\
$^{5}$ South African Astronomical Observatory, P.O.Box 9, 7935     
        Observatory, South Africa \\
$^{6}$  Astronomy Department, University of Cape Town, 7701 Rondebosch, 
        South Africa \\
$^{7}$ NASSP, Department of Mathematics and Applied Mathematics, 
        University of Cape Town, 7701 Rondebosch, South Africa \\        
$^{8}$  Research School of Astronomy \& Astrophysics, Mount Stromlo Observatory,
        Australian National University, 
        Cotter Road, \\
        Weston ACT 2611, Australia \\
$^{9}$  SUPA, School of Physics, University of Edinburgh, IfA, 
        Blackford Hill, Edinburgh EH9 3HJ, United Kingdom \\
$^{10}$  Astrophysics Group, School of Physical and Geographical Sciences, Keele 
        University, Staffordshire ST5 5BG, United Kingdom \\
$^{11}$ Instituut voor Sterrenkunde, KU Leuven, Celestijnenlaan 200D, 
        3001 Leuven, Belgium \\
%$^{10}$  Sterrewacht Leiden, Niels Bohrweg 2, 2333 RA Leiden, The Netherlands \\
%$^{11}$ Institut d'Astrophysique de Paris, CNRS, 98bis Boulevard Arago, 75014 Paris, France \\
        %$^{11}$ Universit\'{e} Paris 6, 98bis Bd Arago, 75014 Paris, France \\
%$^{13}$ Astronomical Institute ``Anton Pannekoek'', University of Amsterdam, 
%        Kruislaan 403, 1098 SJ, Amsterdam, \\
%        The Netherlands \\
             }

\date{Accepted. Received; in original form }
\pagerange{\pageref{firstpage}--\pageref{lastpage}} \pubyear{2007}

\maketitle
\label{firstpage}

\begin{abstract}
 We have observed five carbon-rich AGB stars in the Fornax dwarf spheroidal
(dSph) galaxy, using the Infrared Spectrometer on board the Spitzer Space
Telescope.  The stars were selected from a near-infrared survey of Fornax
and include the three reddest stars, with presumably the highest mass-loss rates,
in that galaxy. Such carbon stars probably belong to the intermediate-age
population (2--8\,Gyr old and metallicity of $\rm [Fe/H]\sim-1$) of Fornax. 
The primary aim of this paper is to investigate mass-loss rate, as a function of
luminosity and metallicity, by comparing AGB stars in several galaxies with
different metallicities.  
The spectra of three stars are fitted with a
radiative transfer model. We find that  mass-loss rates of these three stars
are 4--7$\times10^{-6}$\,$M_{\sun}$\,yr$^{-1}$.  
The other two
stars have mass-loss rates below
$1.3\times10^{-6}$\,$M_{\sun}$\,yr$^{-1}$. We find no evidence that these
rates depend on metallicity, although we do suggest that the gas-to-dust
ratio could be higher than at solar metallicity, in the range 240 to 800.
The C$_2$H$_2$ bands are stronger at lower metallicity because of the higher
C/O ratio.  In contrast, the SiC fraction is reduced at low metallicity, due
to low silicon abundance. The total
mass-loss rate from all known carbon-rich AGB stars into the interstellar
medium of this galaxy is of the order of
$2\times10^{-5}$\,$M_{\sun}$\,yr$^{-1}$.  This is much lower than that of
the dwarf irregular galaxy WLM, which has a similar visual luminosity and metallicity.
The difference is attributed to the younger stellar population of WLM. The
suppressed gas-return rate to the ISM accentuates the difference between the
relatively gas-rich dwarf irregular and the gas-poor dwarf spheroidal
galaxies. Our study will be useful to constrain gas and dust recycling
processes in low metallicity galaxies.
\end{abstract}

\begin{keywords}
stars: AGB and post-AGB -- stars: atmospheres -- stars:mass-loss
stars: carbon -- 
\end{keywords}
%
%________________________________________________________________
\large

%\tableofcontents

\section{Introduction}

Stars with low and intermediate initial mass (from 1 to 8\,$M_{\sun}$)
loose their atmospheres towards the end of their lives
(1--10\,Gyr after their birth).  This mass loss is intense; approximately 50--80\,\% of the
stellar mass is lost during the Asymptotic Giant Branch (AGB) phase.
Understanding this process and the composition of the gas lost from AGB
stars is important for understanding both stellar and galactic evolution. 
First, the stellar wind removes material from AGB stars, reducing their mass
and influencing their evolution. Secondly, AGB stars are among the primary
sources of metal-enriched gas and dust grains in the interstellar medium of
galaxies \citep{Maeder92}.  In particular, carbon-rich dust grains are formed
solely in carbon-rich AGB stars \citep{Dwek98}.  Theoretical work has
suggested that AGB mass-loss rates depend on metallicity
\citep{Bowen91, Willson06}, because the stellar wind is triggered by
radiation pressure on dust grains and the dust is made up of
astronomical metals, such as oxygen, carbon, silicon, iron and aluminium. 
Therefore a study of extra-galactic AGB stars is vital, both because it
allows us to study mass loss at low metallicity, and because the distances
to the parent galaxies are known, so that stellar parameters can be derived
with some precision.  A high sensitivity mid-infrared instrument is
required for such a mass-loss study; at present this is provided uniquely by the {\it Spitzer
Space Telescope} \citep{Werner04}.

The Fornax dSph galaxy is the second (after Sagittarius) most luminous dSph
galaxy known in the Local Group.  The metallicity of red giants has been
measured to be as low as [Fe/H]\footnote{Here we use the chemical abundance
ratio of any element $X$, [$X$/H] = log ($X$/H)-log($X$/H)$_{\sun}$ where
log (H)=12 is the abundance of hydrogen by number of atoms. The metallicity
of the galaxy is represented by iron abundance.}=$-0.7$ and $-1.5$
with a mean value of $-1.0\pm0.3$
\citep{Tolstoy01}. This value is lower than that found in the LMC 
\citep[$-0.4$; median value of red giants in the bar; ][]{Cole05}
and the SMC \citep[$-0.5$ or $-0.73$; ][]{Smith89, vandenBergh99}.  The metallicity of AGB stars
has not yet been determined in these galaxies, except for one AGB star in the SMC,
which has been estimated at $-1.0\pm0.35$ \citep{deLaverny06}. This is consistent,
within the error, with the values quoted above for red giants.  Thus, the Fornax dSph is an
ideal target to explore the lower metallicity range. Fornax has globular
clusters; their ages range from 7 to 14 Gyrs
\citep{Strader03}.  \citet{Battaglia06} found three distinct periods of star
forming activity in the galaxy. The dominant population, to which the AGB
stars are most likely to belong, is 2 to 8 Gyr old.  There may have experienced
star-forming activity until a few times $10^8$\,years ago, but this young
population is small.  
Fornax is gas poor at the present
time;
\citet{Young99} reported a non-detection of H{\small{I}} gas, although
\citet{Bouchard06} have detected more distant gas possibly
associated with Fornax.

The distance modulus of the
Fornax dSph is quoted as: $(m-M)_0$=20.76\,mag \citep{Demers90},
$(m-M)_0$=20.70$\pm$0.12\,mag \citep{Saviane00} and
$(m-M)_0$=20.74$\pm$0.11\,mag \citep{Gullieuszik07}.
%All of these numbers are consistent within the error.
%\citet{Demers90} estimate is based on the 
%optical observations of Fornax globular cluster 1
%while \citet{Saviane00} uses optical observations of Fornax field stars
%and \citep{Gullieuszik07} refer to infrared observations.
These values are consistent and we adopt 20.76\,mag in this paper.  
The
Fornax dSph galaxy is elliptical in shape and the major axis is longer than
the minor axis by a factor of at least two \citep{Irwin95}.  The thickness
of the galaxy is negligible at its distance. The interstellar reddening
towards Fornax is $E(B-V)\approx0.03$ \citep{Demers02}, and its effect
is negligible in the infrared. 

\cite{Demers79} found red giants in the Fornax dSph galaxy and
\citet{Aaronson80} established that some of these are (AGB) carbon
stars.  \citet{Azzopardi99} claim to have detected 104 carbon rich stars in this
galaxy.  Because of its low metallicity, relatively close distance, and
reasonable number of AGB stars, the Fornax dSph is an ideal galaxy to study
the influence of metallicity on the evolution of AGB stars.

As follow-up to our Spitzer spectroscopic survey of AGB stars in the LMC
\citep{Zijlstra06, Matsuura06} and the SMC \citep{Lagadec07}, we have
undertaken spectral observations of carbon stars in Fornax, using the
Infrared Spectrograph (IRS) \citep{Houck04} on the {\it Spitzer Space
Telescope}.  This study aims to investigate the spectra of the AGB stars and
to ascertain the influence of metallicity on AGB mass-loss rates and on the
composition of molecules and dust grains, down to a metallicity of one tenth
of the solar value.

\section{Targets}

Our target selection is based on a monitoring program with the Infrared
Survey Facility \citep[IRSF; ][]{Glass00, Menzies02} at Sutherland in South
Africa (Menzies et al. in preparation).  Table~\ref{table-targets} lists the
five targets together with their coordinates and cross-identifications.  Photometric measurements of
the targets from the literature are summarised in
Table\,\ref{table-magnitudes}.  The variability of these stars will be
discussed elsewhere (Menzies et al. in preparation), but all of
them show large amplitude periodic changes, and are most likely Mira
variables. This variability will result in spectral changes if the stars are
indeed Miras,  \citep[e.g.][]{Hron98, Busso07}.

\begin{figure}
\centering
\resizebox{\hsize}{!}{\includegraphics*{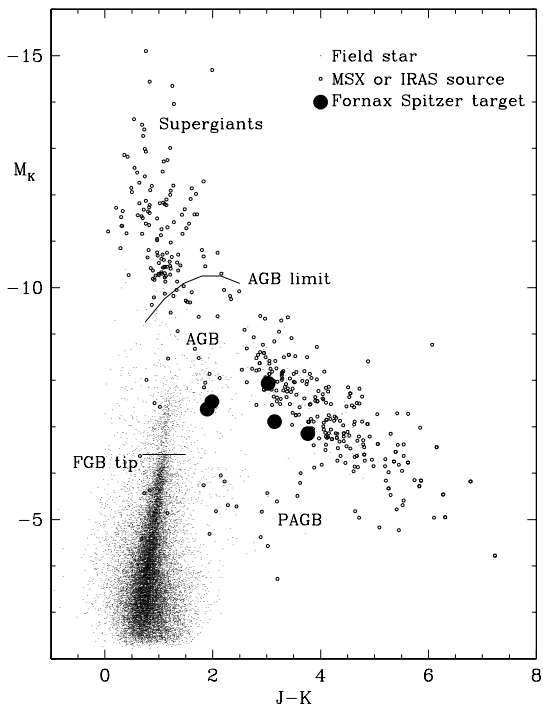}}
\caption{ The absolute $K$-band magnitude ($M_K$), $J-K$ colour 
diagram for the five sources in the 
Fornax dSph galaxy (large filled circles) observed at mid-infrared
wavelengths.  For
comparison a large sample of mid-infrared sources in the LMC from MSX
(Midcourse Space Experiment; small open circles) and field stars from a small area of the LMC bar
(small dots) are also shown --- see \citet{Zijlstra06} and
\citet{Wood07} for details.  AGB stars are confined approximately to
the region below the line marked ``AGB limit''. Stars above this limit are
red supergiants with initial masses above $\sim$8 M$_{\odot}$, or perhaps
foreground stars.  Locations of post-AGB stars (PAGB) and tip of FGB (first giant branch)
are indicated. 
Distance moduli of 20.76 and 18.54 have been assumed for
Fornax and the LMC, respectively.
%Comparision with Marigo et al. (The red tail of carbon stars in the LMC...)
\label{Fig-cm} }
\end{figure}

Fig. \ref{Fig-cm} shows an infrared colour-magnitude diagram of our targets,
where they are compared with LMC stars.  Our sample is located on the
sequence of LMC AGB stars, although the Fornax stars tend to have fainter
$M_K$ magnitudes than do LMC stars with similar $J-K$ colours.

Our three red carbon-rich stars (Fornax 13-23, 12-4, and 3-129) are redder
in $J-K$ than anything described in a recent near-infrared survey of Fornax
\citep{Gullieuszik07} and probably have the thickest shells of any AGB stars
in that galaxy.

%_________________________________________________________________
%_________________________________________________________________

\begin{figure*}
\centering
%\resizebox{\hsize}{!}{\includegraphics*[11,272][560,559]{dss.ps}}
\resizebox{\hsize}{!}{\includegraphics*[50,40][520,340]{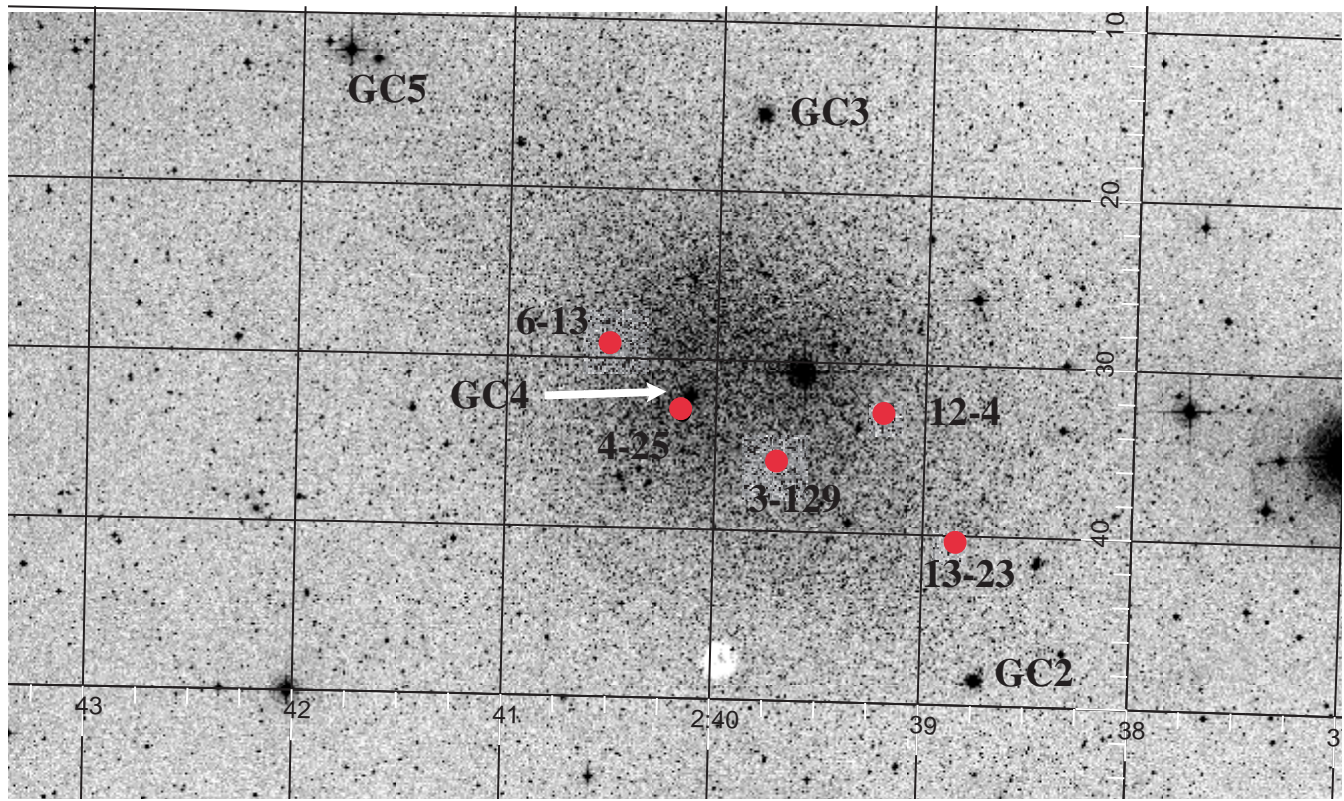}}
\caption{ 
Locations of our target stars within the Fornax dSph galaxy,
superposed on the DSS image.
Globular Clusters (GCs) within the field are also indicated.
Coordinates are in J2000 and one square grid box is 15$\times$10\,arcmin$^2$.
\label{Fig-fornax} }
\end{figure*}

The Fornax dSph galaxy is roughly elliptical in shape, although the
ellipticity varies slightly depending on the distance from the galaxy
centre. All of our targets are located within an ellipse of 0.4\,deg
semi-major axis and ellipticity of 0.3 (see \citet{Battaglia06} their
Fig.\,1; Fig.\,\ref{Fig-fornax} in our work). 
%Within this region, \citet{Battaglia06} describe
%three stellar populations in this region of the Fornax dSph: an ancient
%population ($>$10\,Gyr), an intermediate age one (2 to 8\,Gyr) and a young
%population ($<$1\,Gyr).  THIS ALL SAID IN PREVIOUS SECTION
A colour-magnitude diagram of Fornax
\citep{Battaglia06} shows the `AGB bump', which marks the beginning
of the AGB phase; stars in this bump belong to an intermediate age
population.  Because of the typical age of AGB stars, it is most likely that
our targets also belong to the intermediate age population, which has a
metallicity of [Fe/H]$\sim -1$
%and formed from the same gas distribution 
\citep{Battaglia06}.  Contamination by young ($<$1\,Gyr) carbon stars is
possible, but not likely. Such young carbon stars would have high
luminosities, close to the AGB luminosity limit, which are not found for any
of our sample (Fig.\ref{Fig-cm}). In addition, the population of such young
carbon stars must be small.

\citet{Tolstoy03} found [Fe/H]$\sim -1.6$ among some red
giants from optical high spectral resolution observations, but the
majority were between $-0.9$ and $-1.3$ according to the medium
resolution spectroscopic survey
\citep{Tolstoy01, Battaglia06}.  The median of the `current' metallicity of red
giants is [Fe/H]$\sim -1.0$ according to spectroscopic
observations of over a hundred stars \citep{Pont04}.  Furthermore, a
planetary nebula, which is carbon-rich, shows
[Fe/H]$=-1.13\pm0.18$ \citep{Kniazev07}.  We conclude that [Fe/H]$\sim
-1.0$ is a reasonable assumption for the population we are dealing with.

%The metallicity of carbon stars has been suggested to be as low as 
%[Fe/H]$\sim -1.4$ \citep*{Demers02},
%when these stars were formed.

Two of our targets were identified as carbon stars by \citet{Demers02}.
The other three targets were not included in their sample, because they
selected carbon stars with $1.4<J-K<2.0$.

%_________________________________________________________________
%________________________________________________________________

%_________________________________________________________________
\begin{table*}
 \centering
 \begin{minipage}{160mm}
  \caption{Targets. \label{table-targets}}
% \rotatebox{90}{ 
 \begin{tabular}{lcccllccccccc}
  \hline
Name & Coordinates (J2000) &  & 2MASS designation$^{\dagger}$ & 
 Other names & ref \\ \hline
Fornax 13-23  & 02h38m50.6s & $-$34d40m32.0s & 02385056$-$3440319 & \\
Fornax 12-4    & 02h39m12.3s & $-$34d32m45.0s & 02391232$-$3432450 & \\
Fornax 3-129  & 02h39m41.6s & $-$34d35m56.7s & 02394160$-$3435567 & \\%GHR6299 & 4 \\
Fornax 4-25    & 02h40m10.2s & $-$34d33m21.9s & 02401016$-$3433218 &  V20, C10, DM19 & 1,2,3\\
Fornax 6-13    & 02h40m31.2s & $-$34d28m44.2s & 02403123$-$3428441 &  DM22 & 1,3 \\
\hline
\end{tabular}
\footnotetext{$^{\dagger}$ \citet{Skrutskie06}}
\footnotetext{1: \citet{Demers87} }
\footnotetext{2: \citet{Westerlund87}}
\footnotetext{3: \citet{Demers02} }
%\footnotetext{4: \citet{Gullieuszik07}}
%}
\end{minipage}
\end{table*}
%________________________________________________________________
%_________________________________________________________________
\begin{table*}
 \centering
 \begin{minipage}{160mm}
  \caption{Magnitudes and variability. \label{table-magnitudes}}
% \rotatebox{90}{ 
 \begin{tabular}{lcccccclccccc}
  \hline
  Name  & DI$^{1}$  & & WEL$^{2}$  & &  IRSF$^{3}$ &  & 
2MASS$^{4}$ & & &\\% GHR$^{5}$\\
  & $<V>$ & period & $V$ & $V-B$ &   $<K>$ & $<J-K>$ &
\multicolumn{1}{c}{$J$} & $H$ & $Ks$ \\ %& $J$ & $H$ & $Ks$\\ 
% &&&&&&& 1.235 & 1.662 & 2.159 \\ %& 1.247 & 1.653 & 2.162 \\ 
% &&&&&&&  1594. & 1024. & 666.8 \\ % & 1680 & 1070 & 664 \\ 
 \hline 
  Fornax 13-23  &  &  & &  &13.650$\pm$0.4 & 3.147   & ~~16.106$\pm$0.0939 & 14.525$\pm$0.0527 & 12.879$\pm$0.0287\\
 Fornax 12-4    &  &  & & &  12.823$\pm$0.3 & 3.027   & ~~14.722$\pm$0.0334 &	13.262$\pm$0.0384 & 12.120$\pm$0.0242\\
 Fornax 3-129  &  &  & & &  13.907$\pm$0.4 & 3.764   &  $<$17.663                 & 15.970$\pm$0.2049	& 14.164$\pm$0.0695 \\%&16.135 & 15.573 & 15.410 \\
 Fornax 4-25    & 18.6 & 317.2 &  18.43 & 2.92 &  13.221$\pm$0.2 & 1.984   & ~~14.063$\pm$0.0256 & 13.122$\pm$0.0214 & 12.545$\pm$0.0285 \\
 Fornax 6-13    & & &  & &  13.378$\pm$0.1 & 1.896   & ~~14.745$\pm$0.0433 & 13.689$\pm$0.0456	& 13.072$\pm$0.0365  \\
\hline
\end{tabular}
\footnotetext{
IRSF is a monitoring program providing average magnitudes, while 2MASS
measurements present a single-phase observation}
%\footnotetext{For the 2MASS filters, the central wavelength in $\mu$m and the
%magnitude zeropoint in Jy is listed \citep{Cohen03, vanderBliek00} }
\footnotetext{1: \citet{Demers87} }
\footnotetext{2: \citet{Westerlund87}}
\footnotetext{3: Menzies  et al. in preparation}
\footnotetext{4: \citet{Skrutskie06}}
%\footnotetext{5: \citet{Gullieuszik07}}
%}
\end{minipage}
\end{table*}
%________________________________________________________________

\section{Observations and data reduction}

Our targets were observed with the IRS on board the Spitzer Space
Telescope, using the low resolution mode.  Only the short-low (SL)
module was used with a spectral coverage from 5 to 15 $\mu$m.  This
wavelength range is covered by three segments SL2, SL1 and a bonus
order.  The spectral resolution is in the range from 60 to 127.
The program ID number is 20357.

Observations were carried out on 2006 January 27th to 30th.  The
exposure time per star was 1440~s (60 s$\times$24 cycles) for SL2
(5.2 to 8.7~$\mu$m) and 1680~s (60 s$\times$28 cycles) for SL1
(7.4 to 14.5~$\mu$m (officially), but extends up to 15.0~$\mu$m).  
We used a nearby star, IRAS~F02375$-$3443
(02h39m35.23s, $-$34d30m37.2s), to peak-up on.

%The data were first reduced through the pipeline version S13.2, then spectra were extracted
%using {\it {SMART}}. The flux calibration were {\bf additional information is needed}

The data were processed through the S13.2 and S14.0 version of the
{\it Spitzer} Science Center's pipeline.  The reduction started from
the {\it droop} products which are equivalent to the commonly used
{\it bcd} data, but lack flat-field and stray-cross-light removal
(the latter is only important for bright sources).  Rogue pixels are
first flagged using a campaign mask and then removed using the {\it
irsclean}\footnote{This tool is available from the SSC web site:
http://ssc.spitzer.caltech.edu} tool. Different cycles (repetitions)
were averaged to improve the S/N. Finally, one
dimensional spectra were extracted using a variable extraction window
set at 4 pixels at the middle of each order. The calibration was
performed by dividing the resultant spectrum by that of the
calibration star HR\,6348 (extracted in the same way as the target) and
multiplying by its template (\citet{Cohen03}; Sloan et al. in prep).

\section{Description of the spectra}

\begin{figure*}
\centering
\resizebox{0.9\hsize}{!}{\includegraphics*{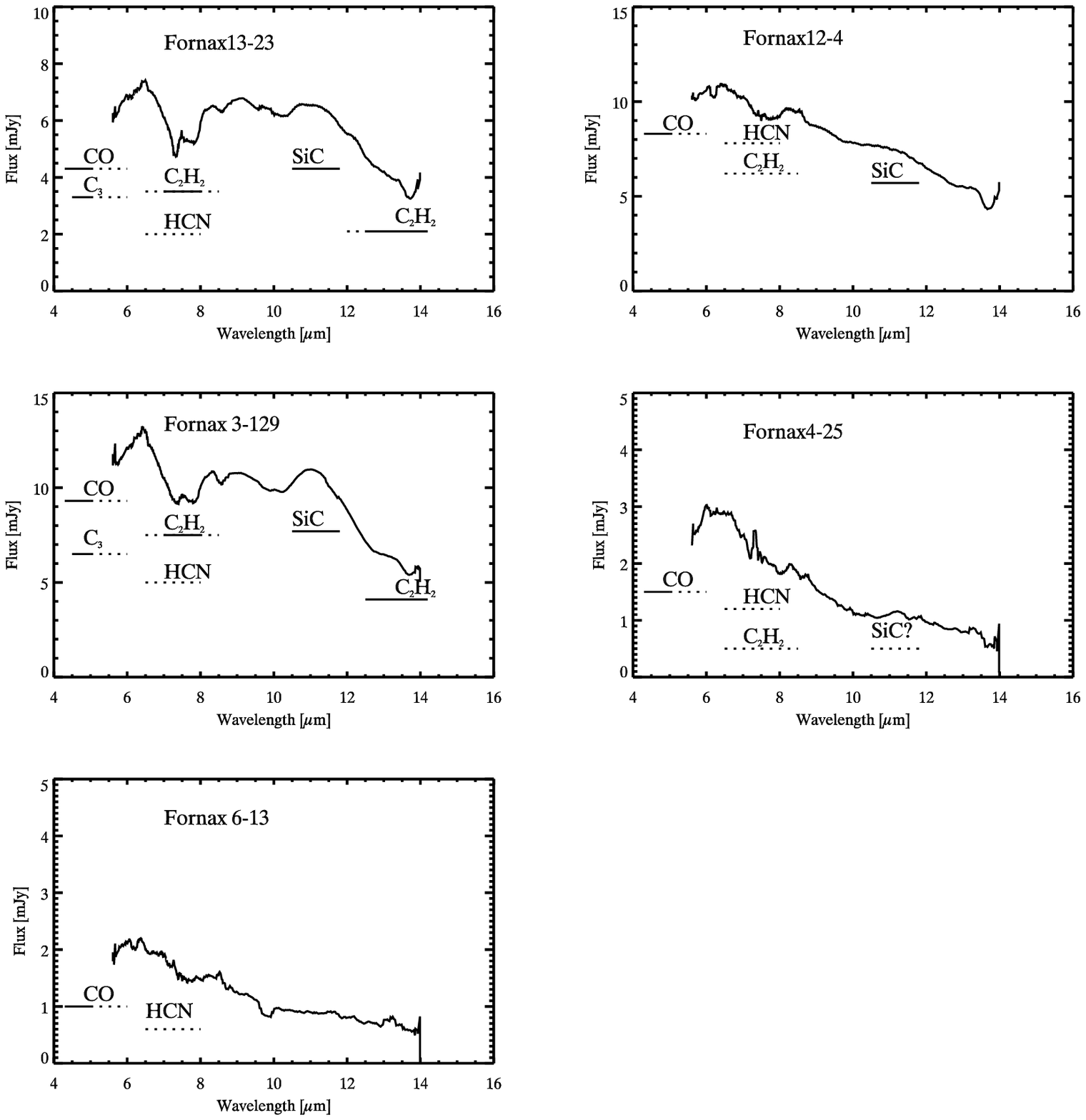}}
\caption{ Spitzer/IRS spectra of carbon-rich stars in the Fornax dSph
galaxy.  Spectra are smoothed by a factor of 2.5 on the wavelength
grid to improve the quality.  The wavelength ranges of the molecular bands are
indicated by lines, and possible identifications of molecular bands
are shown with dotted lines.
\label{Fig-spec1} }
\end{figure*}

Fig.~\ref{Fig-spec1} shows the Spitzer/IRS spectra of the five target stars;
all are carbon-rich.  The target selection (long period variability) was
independent of spectral classification (i.e. O-rich versus C-rich), and this
suggests that high mass loss is only exhibited by carbon stars in this
galaxy.  Absorption features at 7.5 and 13.7\,$\mu$m are due to C$_2$H$_2$.
There may be some contribution from HCN in the absorption feature at 7\,$\mu$m.  The
emission at 11.3\,$\mu$m is a dust excess from SiC.  A dip at 6\,$\mu$m is
part of the 5\,$\mu$m CO fundamental, possibly blended with C$_3$ absorption
bands.  CO molecules will be present in all cool AGB stars.  The
identification of C$_3$ in the Fornax stars is not certain, because the
wavelength coverage of IRS is not favourable for this band, but the presence
of C$_3$ is known for some Galactic (albeit blue) carbon stars
\citep{Jorgensen00,Yamamura97} and therefore it is likely to be present in
these extra-galactic stars.  The SiC excess at 11.3\,$\mu$m is strong in
Fornax 13-23 and Fornax 3-129 and weak in Fornax 12-4.  This feature is
marginally detected in Fornax 4-25 and is not obvious in Fornax 6-13.  The
feature at 8.5\,$\mu$m is caused by the band gap.

Fornax\,6-13 does not show any obvious feature in its IRS spectrum.
However, its near-infrared colour, $J-K>1.4$, suggests
that this star is carbon-rich \citep{Cioni01}, and this is confirmed by
near-infrared spectra (Groenewegen et al. in preparation).
\citet{Demers02} also spectroscopically identified Fornax 4-25 and
Fornax 6-13 as carbon stars.  
%Despite no strong features in their IRS
%spectra, these two stars are justified as carbon stars by other works.

%_________________________________________________________________
%_________________________________________________________________

\subsection{Band strengths}

%_________________________________________________________________
\begin{table}
 \centering
% \begin{minipage}{160mm}
  \caption{ Equivalent width of molecular bands ($W_{7.5}$ and $W_{13.7}$)
  for 7.5 and 13.5\,$\mu$m C$_2$H$_2$, SiC dust emission strength (above the 
continuum)
  with respect to the `continuum' ($R_{\rm {SiC}}$),
  infrared fluxes ($f$) at 6.4 and 9.3\,$\mu$m, respectively and the [6.4]$-$ [9.3] colour; 
  definitions are given by \citet{Zijlstra06}.
  \label{table-spitzer}}
% \rotatebox{90}{ 
 \begin{tabular}{lrrrrrrr}
  \hline
Name & $W_{7.5}$ & $R_{\rm {SiC}}$ & $W_{13.7}$ & $f_{6.4}$ & $f_{9.3}$ & [6.4]$-$[9.3] \\
 & $\mu$m & & $\mu$m & mJy & mJy & mag \\ \hline
For 13-23  & 0.25 &   0.14     &    0.022 &      7.2 &   6.7 &     0.73 \\
For 12-4    & 0.08 &    0.04    &    0.058 &     10.8 &  8.4  &    0.55 \\
For 3-129  & 0.22 &    0.15    &    0.029 &    12.7 & 10.6  &   0.62 \\
For 4-25    & 0.15 &    0.03    &    0.019 &      2.9 &    1.4   &  0.04 \\
For 6-13    & 0.11 &   0.01     &    0.026 &      2.1 &    1.2   &  0.20 \\
%
%Fornax13-23       0.2540064              0.1407140       0.0218080              0.0071997      10.3180399       0.0066672       9.5899353
%Fornax12-4       0.0762908              0.0436796       0.0580508              0.0107777       9.8800077       0.0084438       9.3334494
%Fornax3-129       0.2155217              0.1546620       0.0287229              0.0126949       9.7022467       0.0106065       9.0858574
%Fornax4-25       0.1505050              0.0320306       0.0191675              0.0028722      11.3157921       0.0014136      11.2739897
%Fornax6-13       0.1124893              0.0126348       0.0260708              0.0020922      11.6598177       0.0011877      11.4630079
%
\hline
\end{tabular}
%}
%\end{minipage}
\end{table}
%________________________________________________________________

\begin{figure}
\centering
\resizebox{\hsize}{!}{\includegraphics*{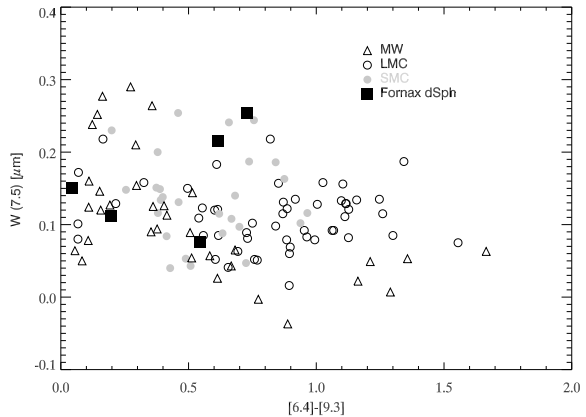}}
\caption{  
The equivalent width of 7.5\,$\mu$m C$_2$H$_2$ as a function of 
infrared colour
[6.4]$-$[9.3]. Symbols show the host galaxies,
MW representing the Milky Way.
\label{Fig-c69-eq7} }
\end{figure}

\begin{figure}
\centering
\resizebox{\hsize}{!}{\includegraphics*{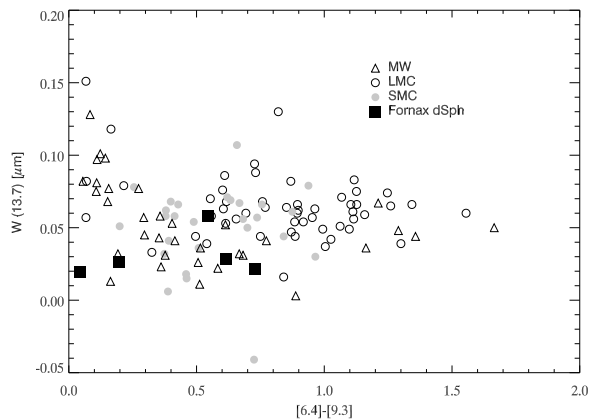}}
\caption{  The same as Fig.\,\ref{Fig-c69-eq7} but for 
the equivalent width of 13.7\,$\mu$m C$_2$H$_2$.
\label{Fig-c69-eq13} }
\end{figure}

We measured the equivalent width of the molecular bands and the strength of
the SiC dust emission, so as to evaluate the metallicity dependence of these
features.  The equivalent width of the C$_2$H$_2$ molecular bands and SiC
are measured following the method of \citet{Zijlstra06}.
Figs.\,\ref{Fig-c69-eq7} and \ref{Fig-c69-eq13} show 7.5 and 13.7\,$\mu$m
C$_2$H$_2$ equivalent widths as a function of infrared colour [6.4]$-$[9.3]. 
The [6.4] and [9.3] values are calculated from the Spitzer/IRS spectra, and
the definition of these magnitudes is also given by \citet{Zijlstra06}; the
bands being selected so as to avoid major molecular features.  [6.4]$-$[9.3]
is a measure of the stellar temperature for blue stars, but of circumstellar
excess for red stars.  Samples in our Galaxy (Milky Way), the LMC and the
SMC are also plotted in the figures \citep{Sloan03, Sloan06, Zijlstra06,
Lagadec07}.  As seen in the spectra and Table\,\ref{table-spitzer}, Fornax 13-23 and 3-129 have a
large equivalent width at 7.5\,$\mu$m (W(7.5)=0.2--0.25\,$\mu$m) with
respect to the infrared colour [6.4]$-$[9.3]. Here W is the acronym of `equivalent Width'. 
These band strengths are some
of the largest found in the range $0.5<$[6.4]$-$[9.3]$<1.0$, within the
samples from various galaxies.  
A high value of W(7.5), with respect to
infrared colour, is found for stars in Fornax and the SMC, while a low
W(7.5) is found in our Galaxy.  The equivalent width of 13.7\,$\mu$m is more
affected by spike noise because this molecular feature is weak,
and the quality of data in 14.5--15.0~$\mu$m region is poorer.
Nevertheless, Fig.\,\ref{Fig-c69-eq13} shows that
stars in our Galaxy cover the lowest range of 13.7\,$\mu$m
equivalent width, W(13.7) in the range $0.5<$[6.4]$-$[9.3]$<1.0$.  

%{\bf explanation of 7 micron eq without feature}.

\begin{figure}
\centering
\resizebox{\hsize}{!}{\includegraphics*{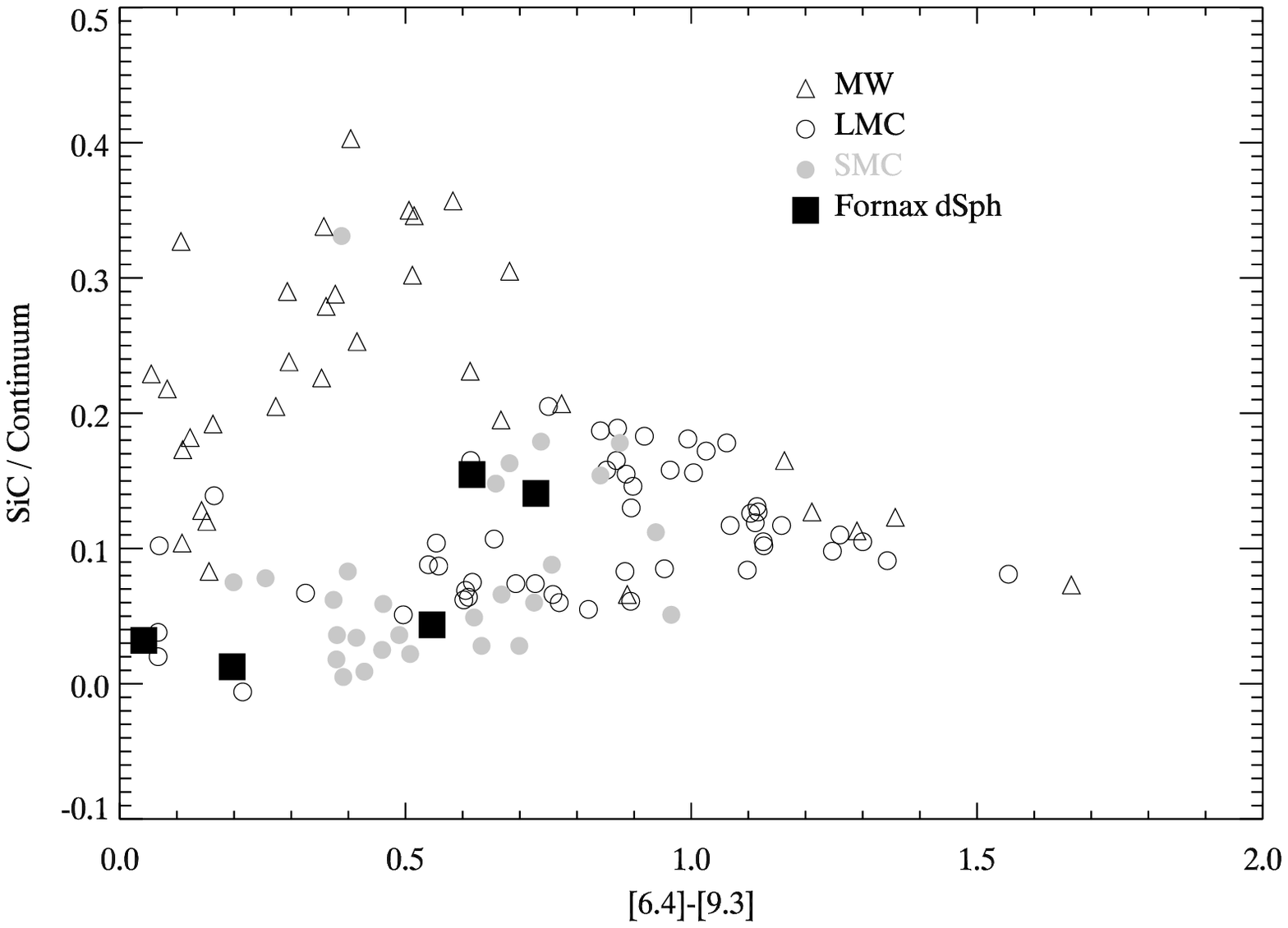}}
\caption{  
The intensity ratio of SiC excess with respect to the pseudo-continuum
is plotted against [6.4]$-$[9.3];
a ratio of 0.0 indicates no SiC excess (continuum only).
\label{Fig-c69-sic} }
\end{figure}

Fig.\,\ref{Fig-c69-sic} shows the ratio of SiC excess with respect to
the pseudo-continuum \citep{Zijlstra06}; larger ratios indicate larger SiC
excess.  A careful positioning of the `continuum' would be required to
measure the SiC band strength precisely \citep{Thompson06}, but we argue
that this ratio represents a good estimate of the approximate SiC strength. 
The ratio is plotted against the infrared colour [6.4]$-$[9.3] in
Fig.\,\ref{Fig-c69-sic}. A large SiC excess is found for relatively blue
(IR-colour $0.0<$[6.4]$-$[9.3]$<0.5$) Galactic stars. Stars in the SMC and
Fornax have the lowest SiC excess ratio among any of the samples. Two stars
in Fornax show a SiC to continuum ratio of $\sim$0.15 at
$0.5<$[6.4]$-$[9.3]$<0.8$, much less than found for Galactic stars with
similar colour.

%_________________________________________________________________
%_________________________________________________________________
%_________________________________________________________________
%_________________________________________________________________
%_________________________________________________________________
%_________________________________________________________________

\section{Mass-loss rate}
 Using flux calibrated Spitzer spectra together with infrared photometric data,
 we derive mass-loss rates for the carbon-rich AGB stars.
%_________________________________________________________________
\begin{figure}
\centering
\resizebox{\hsize}{!}{\includegraphics*{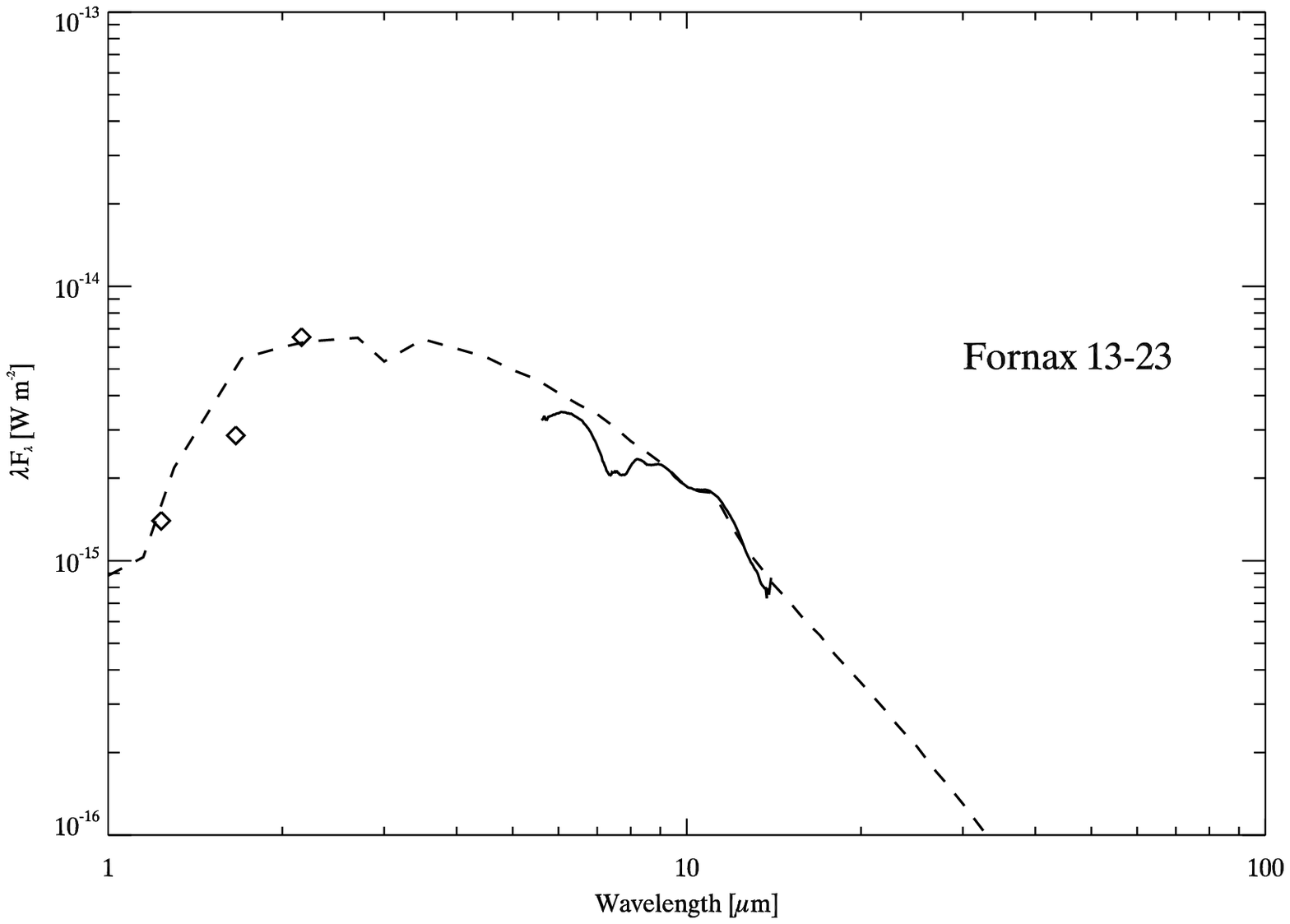}}
\resizebox{\hsize}{!}{\includegraphics*{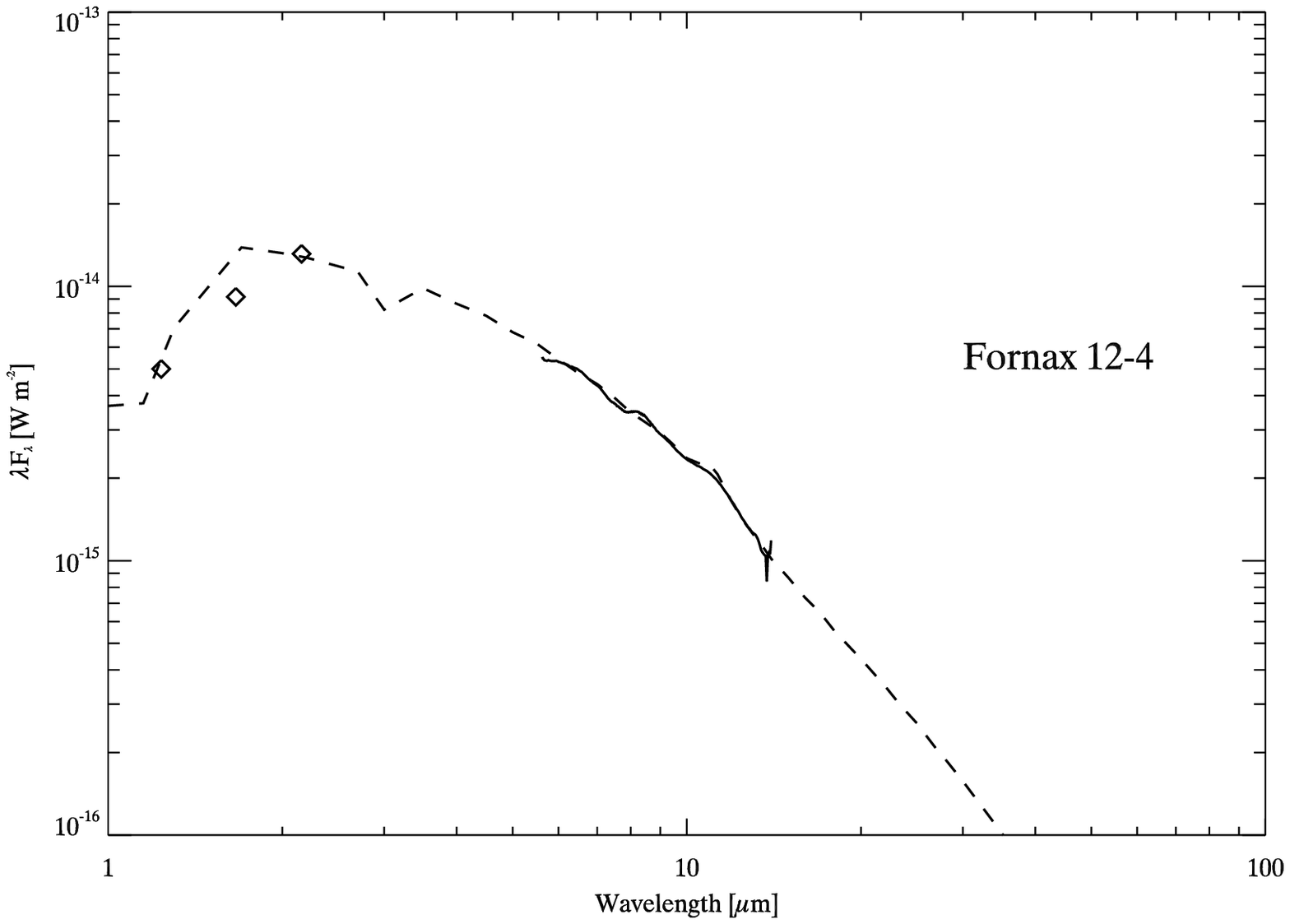}}
\resizebox{\hsize}{!}{\includegraphics*{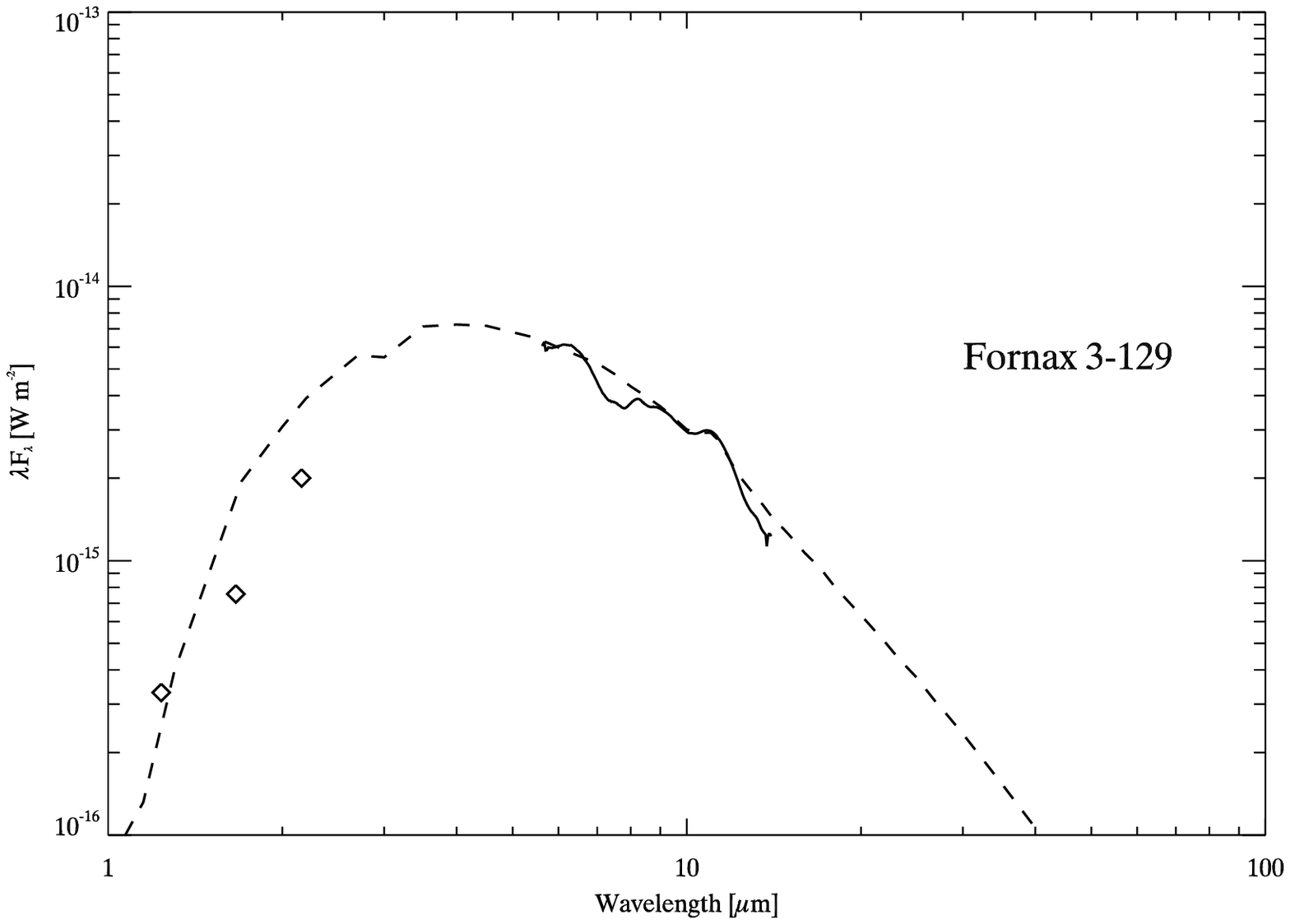}}
\caption{  
SED and fits with {\small DUSTY} radiative transfer model.
\label{fig-dusty} }
\end{figure}
%_________________________________________________________________

%_________________________________________________________________
\begin{figure}
\centering
\resizebox{\hsize}{!}{\includegraphics*{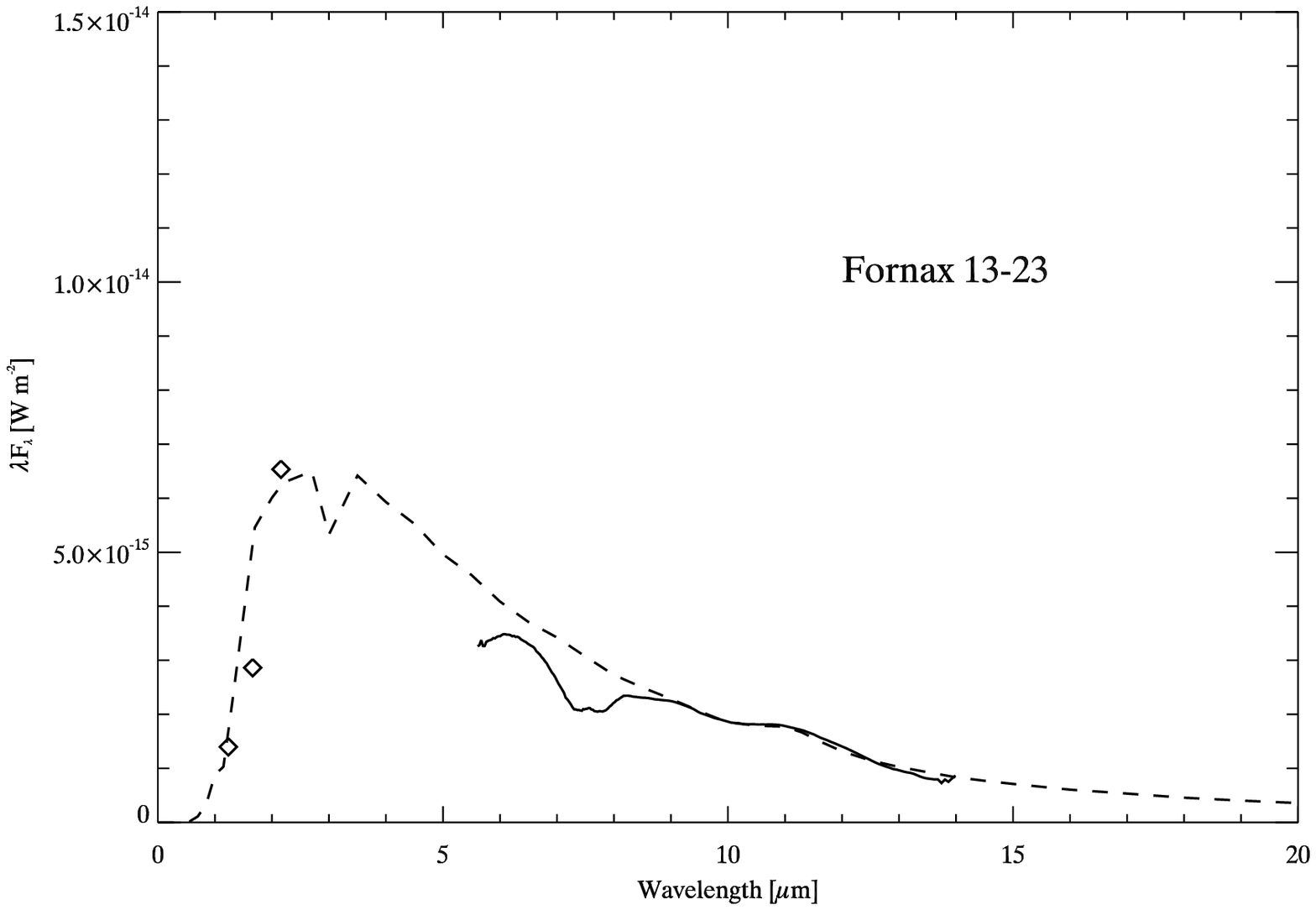}}
\resizebox{\hsize}{!}{\includegraphics*{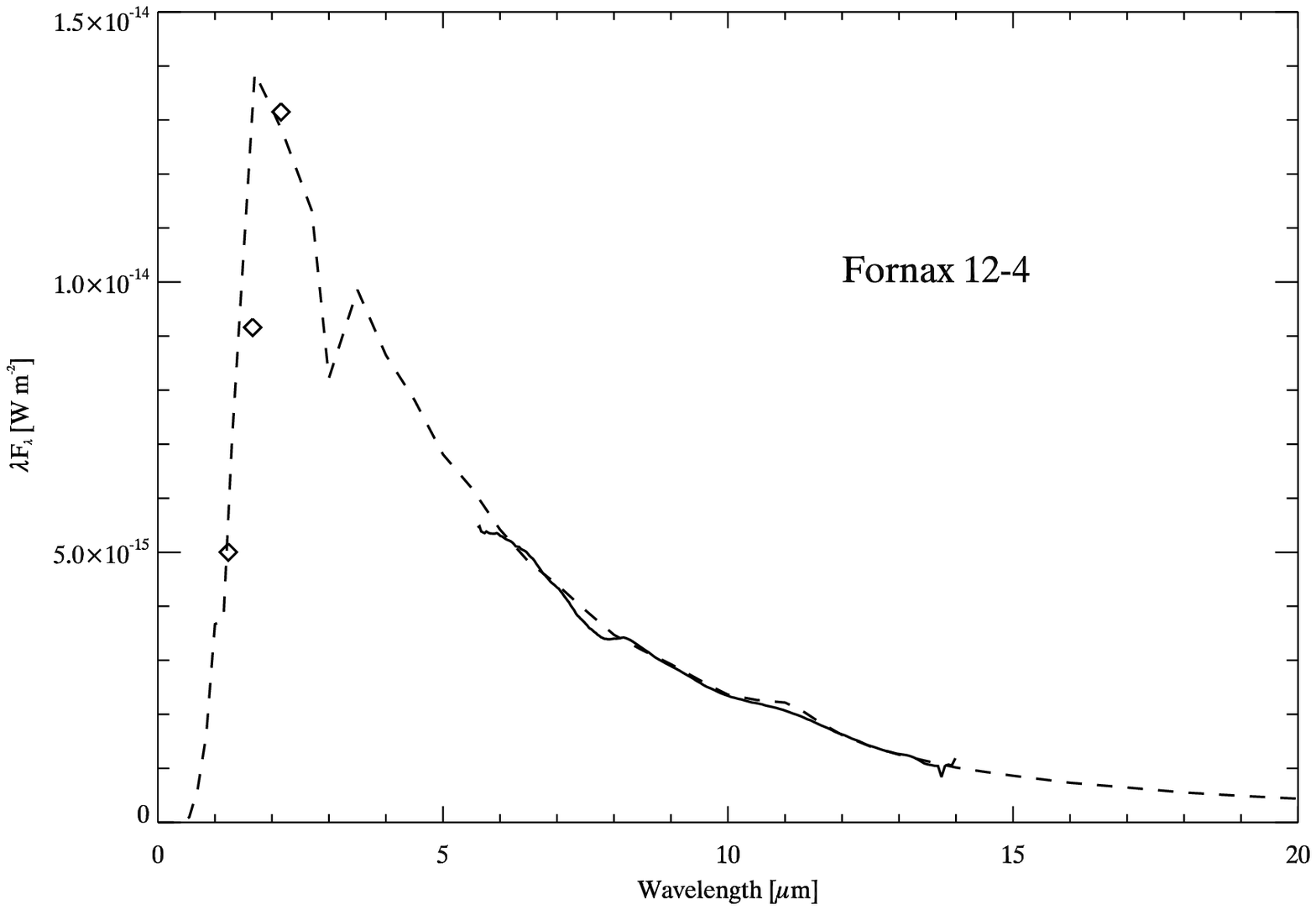}}
\resizebox{\hsize}{!}{\includegraphics*{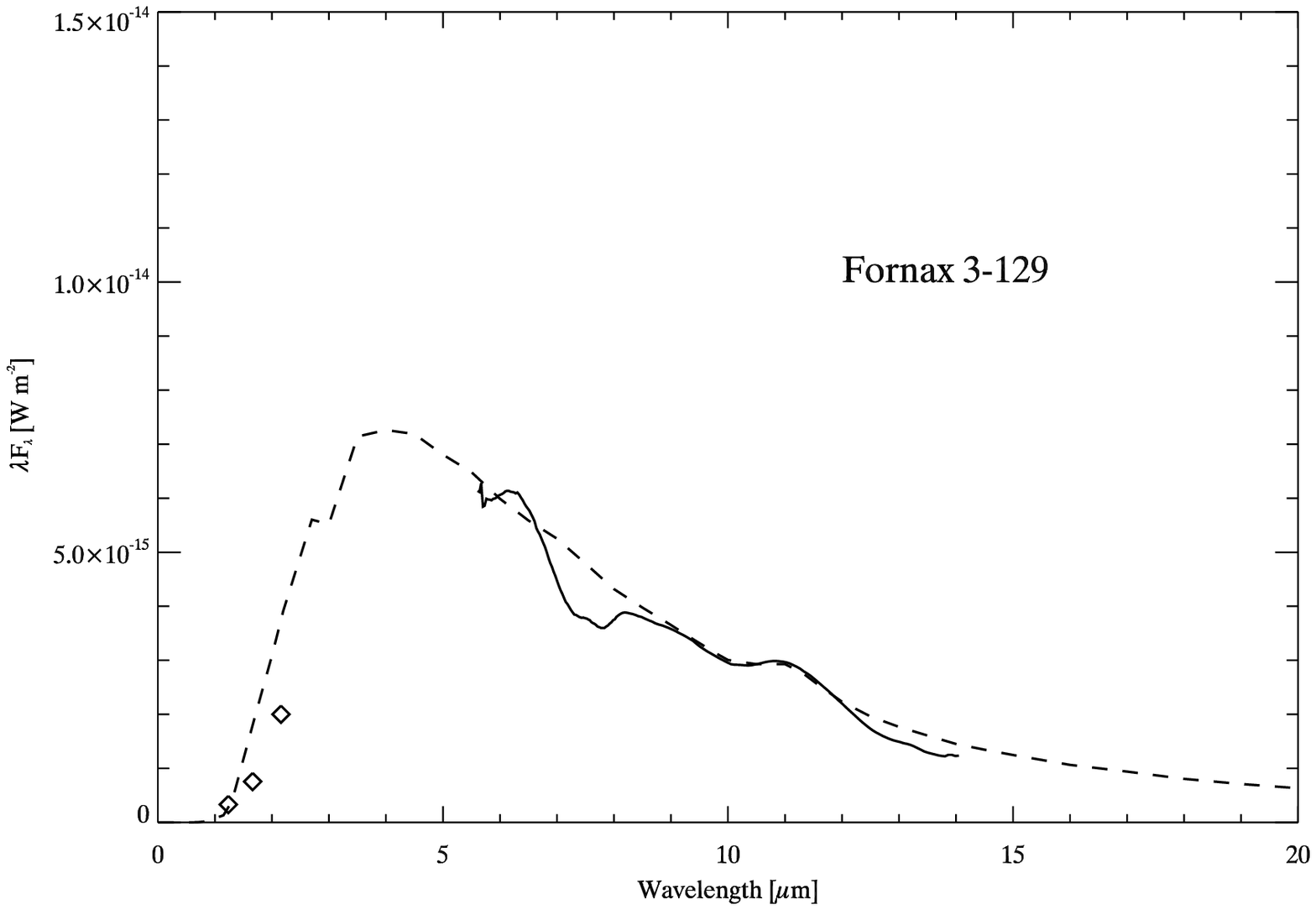}}
\caption{  
SED and fits with {\small DUSTY}, but with a linear scale to emphasize 
the SiC dust feature and quality of the fit.
\label{fig-dusty-lin} }
\end{figure}
%_________________________________________________________________

%_________________________________________________________________
\begin{table*}
 \centering
 \begin{minipage}{160mm}
  \caption{ Luminosity and mass-loss rate. \label{table-dusty}}
% \rotatebox{90}{ 
 \begin{tabular}{lllrccccccc}
  \hline
Name & $M_{\rm{bol}}$ & $L$  & $\tau_{0.55}$ &$\dot M^1$ & $v$ & $\dot M_{\rm {d}}^2$ \\ \hline
% for 1000 Lum 
%Fornax 13-23 &  $-$4.92 &   7389 & 3.00 & 5.22$\times10^{-8}$ & 21.3 & 5.43474e-09\\
%Fornax 12-4   &  $-$5.54 &  13012 & 2.45 & 3.72$\times10^{-8}$ & 23.1 & 3.81259e-09\\
%Fornax 3-129 &  $-$4.97 &   7748  & 8.50 & 8.47$\times10^{-8}$ & 17.3 & 2.92156e-08\\
For 13-23 &  $-$4.92 &   ~~7389 & 4.00 & ~~4.2$\times10^{-6}$ & 20 & ~~2.2$\times10^{-6}$\\
For 12-4   &  $-$5.54 &  13012 & 2.45 & ~~4.5$\times10^{-6}$ & 25 & ~~2.4$\times10^{-6}$\\
For 3-129 &  $-$4.97 &   ~~7748  & 8.50 & ~~6.9$\times10^{-6}$ & 16 & \\
%Fornax 4-25 & $-4.59$$^{\dagger}$  & 5621$^{\dagger}$ &&&&  \\
For 6-13 & $-4.69$$^{\dagger}$  & ~~5721$^{\dagger}$ & $<$1~~~~~& $<$1.3$\times10^{-6}$&&  \\
\hline
\end{tabular}
\footnotetext{ $M_{\rm{bol}}$ : bolometric luminosity (mag) estimated from
{\small DUSTY} fitting.} 
\footnotetext{ $L $ : luminosity ($L_{\sun}$)
estimated from {\small DUSTY} fitting.}
\footnotetext{$^{\dagger}$ : spline fitting to the SED is used instead of 
{\small DUSTY}.}  
\footnotetext{ $\tau_{0.55}$ :
optical depth at 0.55\,$\mu$m for {\small DUSTY} model fitting }
\footnotetext{ $\dot M^1$ : gas mass-loss rate ($M_{\sun}$\,yr$^{-1}$)
estimated from {\small DUSTY} fitting } 
\footnotetext{ $v$ : terminal
velocity (km\,s$^{-1}$) estimated from {\small DUSTY} fitting}
\footnotetext{ $\dot M_{\rm {d}}^2$ : gas mass-loss rate estimated
from near- and mid-infrared colour method, assuming a gas-to-dust
ratio of 200 \citep{Lagadec07b}.  The $K-[11]$ colour is used. The
values quoted in the table are based on method (2) in their paper.}
%}
\end{minipage}
\end{table*}
%________________________________________________________________

\subsection{Modelling the spectral energy distribution}

Spectral energy distributions (SEDs) are modeled using the radiative
transfer code {\small DUSTY} \citep{Ivezic97}.  The quality of these 
SED fits with the observations are illustrated in Fig.\,\ref{fig-dusty}.

A spherical geometry is assumed for the circumstellar envelope, and the
circumstellar shells are filled with material from the radiatively driven
wind, which is moving at a constant velocity.  The photospheric spectrum of
the central star is calculated with a hydrostatic model including molecular
opacities \citep{Loidl01, Groenewegen07}.  An effective temperature of
2800\,K is adopted.  We also tried a 2650\,K spectrum which 
often worked well for stars in the LMC and SMC
\citep{Groenewegen07}, but this lower temperature did not result in
a good fit for the Fornax spectra.  This may be a consequence of the
difference between the samples from the various galaxies; the LMC and SMC
stars have redder $J-K$ colours (Fig.\,\ref{Fig-cm}).  As details of carbon
to oxygen ratios are unknown, a C/O ratio of 1.1 is assumed in the
hydrostatic model. Although this is typical of values found for Galactic
carbon stars, at lower metallicity the C/O ratio may well be higher
\citep{Matsuura02, Matsuura05}.  The assumed C/O ratio affects most strongly
the depth of molecular bands, and we minimize its impact by concentrating on
a relatively featureless wavelength range of the spectra for this particular
exercise.  Furthermore, the elemental abundances (except for carbon) used by
{\small DUSTY} modelling are the solar values \citep{Anders89}, which are also
inappropriate for stars in Fornax.  In spite of these limitations, we
prefer the hydrostatic model spectra over simple blackbodies (BBs).  We
compared SED fits using  blackbody curves 
the star and hydrostatic model spectra as a heating input of {\small DUSTY}.  The simple blackbody
requires a lower optical depth to reproduce the SED.
For example, Fornax 13-23 and Fornax 3-129 are fitted
with $\tau_{0.55}=$3.0, and 8.25 with BB input, respectively.
However, the models with BB input have difficult in fitting near-infrared spectra.
The low optical depth required by the BB model is probably a consequence of
the lack of molecular absorption, thus the more efficient heating of grains especially
at shorter wavelength.

The temperature at the inner radius of the dust shell is assumed to be
1000\,K \citep{Groenewegen07}.  Dust grains are a mixture of amorphous
carbon \citep{Hanner88} and SiC \citep{Pegourie88}.  The fractions which
give the best fit are 5\,\% SiC and 95\,\% amorphous carbon.  We also
attempted to fit the observed SiC bands with 3\,\% and 10\,\% SiC mixtures,
but the former does not create any SiC bands at 11\,$\mu$m, while the latter
produces SiC bands that are too strong to fit the spectra of our targets. 
The grain size distribution is assumed to be $a=0.0005$ to 0.25\,$\mu$m
according to a power-law distribution of the form $n(a) \propto a^{-q}$ with
$q=3.5$ \citep{Mathis77}.  A gas-to-dust ratio of 200 is the default of
{\small DUSTY}. Although we note that the ratio might be higher at lower
metallicity \citep{vanLoon00}, no dependence has yet been measured.  This
ratio is important when the optical depth is converted to the mass-loss rate.

The outer radius of the shell is fixed at $10^3$ times the inner
radius of the circumstellar shell.  Changing the outer radius has 
minor effects on the SED \citep{vanLoon03}.  The {\small DUSTY}
model fit produces the terminal outflow velocity and mass-loss rate for
a luminosity of $10^4 L_{\sun}$.  As we derive the luminosity by
scaling the emerging spectrum from {\small DUSTY}, we recalculate the
dust mass-loss rate accordingly.

The {\small DUSTY} model itself has an uncertainty of 30\,\%
in mass-loss rate and expansion velocity, as noted in the {\small DUSTY}
manual.

\subsection{Results of radiative transfer code fitting}

Figs. \ref{fig-dusty} and \ref{fig-dusty-lin} shows the {\small{DUSTY}} model fits to the
spectra and the parameters used for these fits are listed in
Table\,\ref{table-dusty}.  We use three spectra which show clear SiC
excesses to fit the 11.3\,$\mu$m SiC excess and the general shape of
the SED.  We attempted to fit the other two spectra with negligible SiC excess
 (Fornax 4-25 and Fornax 6-13) and found that the optical depth at 0.55\,$\mu$m was
$\tau_{0.55} < 1$.  The Fornax 4-25 and 6-13 spectra seem to be photospheric
in origin, with little circumstellar contribution to the infrared. 
Integrating over the $JHK$ photometry and the Spitzer spectra of Fornax
6-13, using spline fitting \citep{Zijlstra06}, we obtained a luminosity of
5721\,$L_{\sun}$; {\small{DUSTY}} fitting gave an upper limit to the
mass-loss rate of $< 1.3\times10^{-6}$\,$M_{\sun}$\,yr$^{-1}$.  The same
spline fitting is adopted for Fornax 4-25, using $VJHK$ and Spitzer spectra,
resulting in a luminosity of 5621\,$L_{\sun}$, but with lower accuracy than
for the other targets.  An upper limit to the
mass-loss rate is $< 1.3\times10^{-6}$\,$M_{\sun}$\,yr$^{-1}$.
These photometric observations are not simultaneous
with the Spitzer measurements, in contrast to the study by
\citet{Groenewegen07}, increasing the uncertainty in the instantaneous
luminosities.

There are some features which the {\small DUSTY} models do not fit well, in
particular the 7.5\,$\mu$m and 13.7\,$\mu$m absorptions due to C$_2$H$_2$. 
A similar problem was also found in fitting some LMC and SMC AGB spectra
\citep{Groenewegen07}. The C$_2$H$_2$ molecular bands are included in the
model for the stellar atmosphere, which is used as an input to {\small
DUSTY} models, but the observed bands are stronger than predicted.  As
discussed later, the parameters used in the models are appropriate for
Galactic stars, and the metallicity dependence of the molecular band
strengths may well contribute to this problem.  There could be additional
effects, such as some C$_2$H$_2$ originating from the extended atmosphere
which is lifted by the pulsation above the photosphere
\citep[e.g.][]{Hron98, Ohnaka07}. Circumstellar molecular absorption
\citep[e.g.][]{Aoki99,Matsuura06} is also not included in {\small DUSTY}.

The mass-loss rates for the three reddest stars are of the order of $5
\times 10^{-6}$\,$M_{\sun} yr^{-1}$.  This would be only a moderate mass-loss rate
for Galactic carbon stars.  The comparison assumes that the gas-to-dust
ratio in the Fornax dSph galaxy is the same as in our Galaxy.

The bolometric luminosities  estimated from fitting a
radiative transfer model  are comparable to other estimates,
such as those from near-infrared, $J-K$, colours \citep{Lagadec07b}.
The values range between $\log\,L\,[L_{\sun}]$=3.9 and 4.1, which is
comparable to the mean luminosities found for carbon stars in the LMC,
but slightly higher than those found for SMC stars \citep{Zijlstra06,
Buchanan06, Lagadec07}.

\begin{figure}
\centering
%\resizebox{\hsize}{!}{\includegraphics*[130,460][430,730]{lum_massloss2.ps}}
%\resizebox{\hsize}{!}{\includegraphics*[83,193][509,595]{lum_massloss2.ps}}
\resizebox{\hsize}{!}{\includegraphics*[52,23][475,418]{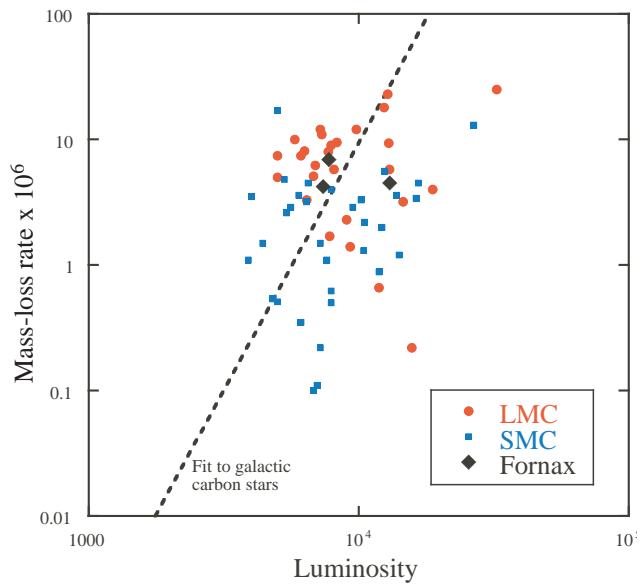}}
\caption{  
Mass-loss rate ($M_{\sun}$\,yr$^{-1}$) as a function of luminosity ($L_{\sun}$)
for our sample and for the LMC and SMC samples \citep{Groenewegen07}.
Fornax stars show high  mass-loss rates for their  luminosities.
The dotted line is the fit to the luminosity vs mass-loss rate relation
for Galactic carbon stars \citep{Groenewegen98}.
\label{lum_massloss} }
\end{figure}

Fig.\,\ref{lum_massloss} shows the derived mass-loss rate as a function of
luminosity and a comparison with values for LMC and SMC stars
\citep{Groenewegen07}. 
Little metallicity dependence of the mass-loss rate is found,
at least for the dust mass-loss rate.
The Fornax mass-loss rates are at the upper end of the SMC
mass-loss rates at a given luminosity.  However, we are unable to fit the
Fornax stars with lower mass-loss rates (Fornax 4-25 \& Fornax 6-13), so
there is an obvious selection effect.  LMC stars appear to reach a higher
mass-loss rate than the SMC and Fornax stars at a given luminosity, as noted
by \citet{Groenewegen07}.

The assumptions made in deriving the total mass-loss rates are
the expansion velocity of the circumstellar shell and the conversion factor
from dust mass-loss rate to gas mass-loss rate.

\citet{Groenewegen07} assumed an expansion velocity of 10\,km\,s$^{-1}$
for all the stars in their study. The velocity is derived from {\small
DUSTY} fitting in our work and we find that the Fornax stars
have expansion velocities in the 16 to 25\,km\,s$^{-1}$ range. This gives us
larger mass-loss rates for the Fornax stars by a factor of up to 2.5 over
those obtained from the \citet{Groenewegen07} assumption. 
Nevertheless, in the log scale of Fig.\,\ref{lum_massloss}, this is a minor
effect.

The gas-to-dust mass ratio is assumed to be 200 in both this work and by
\citet{Groenewegen07}. This ratio may have a metallicity dependence, and
\citet{vanLoon00} estimates $ log\,\psi \propto-1.0\times$[Z/H] for
carbon stars, where $\psi$ is the gas-to-dust ratio.  If this is correct, the gas-to-dust
ratio in Fornax will be about twice what it is in the SMC and the Fornax
stars will have higher gas mass-loss rates than derived above.  We discuss
this further below (Sect.\,\ref{metal-dependence}).

\begin{figure}
\centering
\resizebox{\hsize}{!}{\includegraphics*[158,432][426,685]{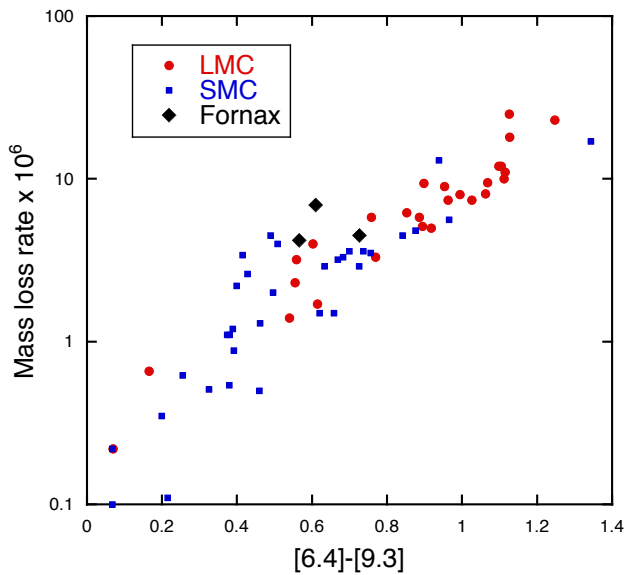}}
%\resizebox{\hsize}{!}{\includegraphics*{color_massloss.ps}}
\caption{  Mass-loss rate  as a function of infrared colour [6.4]$-$[9.3]
\label{color_massloss} }
\end{figure}

Fig.\,\ref{color_massloss} shows the mass-loss rate as a function of
infrared colour.   The wavelength bands are defined so as to avoid major
molecular or dust features, thus the colour measures the `continuum'
spectrum reasonably well.  In Fig.\,\ref{color_massloss}, a correlation is
found between the infrared colour and the mass-loss rate. As mentioned in
\citet{Groenewegen07}, this colour gives the approximate mass-loss rate with
a factor of two to five uncertainty, depending on the colour.

%A comparison of Fig.\,\ref{lum_massloss} with Fig.\,\ref{color_massloss}
%shows that the LMC sample contains stars with significantly redder 
%colours than the other galaxies and that many of these have high luminosity.
%The lack of such red stars in Fornax may indicate that the galaxy
%lacks the somewhat younger populations present in the LMC.
%IS THIS TRUE? IT LOOKS TO ME AS IF THE V RED LMC STARS HAVE A RANGE OF
%LUMINOSITY. 

%_________________________________________________________________
%_________________________________________________________________

%_________________________________________________________________
%_________________________________________________________________

\subsection{A comparison with IR colour method}

\citet{Lagadec07b} estimated the dust mass-loss rate of two of the sources
in our sample, using the infrared colour index $K-[11]$.  Their dust
mass-loss rates have been converted to gas mass-loss rates assuming a
gas-to-dust ratio of 200 and are listed in Table\,\ref{table-dusty}; they are
consistent with our estimates to within a factor of two.  At least some of
the difference will be due to the input stellar spectra, i.e., our work uses
a hydrostatic model while
\citet{Lagadec07b} assume a blackbody.  Nevertheless, the 
consistency shows that the infrared colour method can be used to estimate
the mass-loss rates, as found for Galactic AGB stars
\citep{Whitelock94,Lebertre97}.

The mid-infrared colour also shows a correlation with mass-loss rate
(Fig.\,\ref{color_massloss}), especially when using colours that minimize
the influence of strong molecular bands.  Note that unlike some colour
measurements, e.g. $K-[12]$, our values are constructed from spectra and are
therefore simultaneous and not affected by variability.

\section{Discussion}

\subsection{Luminosity of carbon stars}

The carbon stars in the Fornax dSph galaxy tend to have fainter absolute $K$
magnitudes at a given $J-K$ colour (Fig.\,\ref{Fig-cm}) than LMC carbon
stars.  \citet{Demers02} pointed out that the mean magnitudes
of carbon-rich stars at the tip of the AGB are $<$$M_K$$>=-7.91$\,mag and
$<$$M_K$$>=-7.88$\,mag for the LMC and SMC, respectively, while the mean
magnitude of 26 carbon stars in Fornax gives
$<$$M_K$$>=-7.68$\,mag.  The difference between the LMC and SMC samples was
also discussed in
\citet[Fig.~2;][]{Lagadec07}.  
\citet{Demers02} concluded that the low $<$$M_K$$>$
is due to the lack of younger (0.3--1~Gyr) carbon stars in Fornax.
% Theoretical work by \citet{Mouhcine02} predicts that
%young carbon stars are about $\sim1$\,mag brighter than older ones.
The population analysis by \citet{Battaglia06} also supports this
interpretation; the star formation rate declined during the period from a
few hundred Gyr to 1\,Gyr ago, so we are unlikely to find young (0.3--1 Gyr)
carbon stars in Fornax.  In other words, AGB stars in the Fornax dSph Galaxy
are, on average, probably older than those found in the LMC and the SMC.
%{\bf evolutionary track}

\subsection{Metallicity dependence of mass-loss rates}\label{metal-dependence}

The ratio of SiC to continuum is smaller in Fornax stars than in our
Galaxy or in the LMC, but comparable to that in the SMC. The
Fornax spectra are well fitted if the mixture of dust grains is 5\,\% SiC
and 95\,\% amorphous carbon, although the error in our estimate is
large. As \citet{Groenewegen07} found, the fraction of
SiC dust grains is lower in carbon stars in the LMC and SMC than those in
our Galaxy (10\,\%; e.g. \citet{Groenewegen95}). Stars in the Fornax dSph
galaxy follow this same trend with metallicity.

The fraction of SiC dust in the grains is probably correlated with the silicon
elemental abundance in the atmospheres of carbon stars.  
%Silicon ($^{28}$Si) is an $\alpha$-process element and type II supernovae are
%its most likely source \citep[e.g][]{Iwamoto99}.  
The temperature in
the helium burning shell does not become high enough for
$\alpha$-captures on magnesium, and the abundance of $^{28}$Si can be
assumed to be unchanged during the AGB phase.  The abundance of silicon has
been measured in two red-giants in the Fornax dSph galaxy
\citep{Shetrone03}. They find silicon abundances in metal poor stars higher
than implied by a simple scaling from the metallicity as represented by [Fe/H].
These two stars have approximately 5 and 25\,\% of the solar silicon
abundance ([Si/H]\,=\,$-1.3$ and $-0.6$), while their iron abundances are
[Fe/H]\,=\,$-1.60$ and $-0.67$.  Thus, the metal-poor star shows a
silicon abundance excess, presumably due to the detailed history of its formation
\citep{Shetrone03}.
%This excess facilitates the formation of SiC. The more metal-rich star
%shows a solar [Si/Fe], and the Si excess appears to be limited to the
%oldest Fornax population. 

A mass fraction of 5\%\ SiC grains and a gas-to-dust ratio of 200 in mass,
implies that 25\%\ of the silicon atoms are in the dust, if the envelope has a solar
metallicity. Here we assume the solar abundance of silicate is 7.55,
on a scale of log N(H)=12. 
For [Si/H]\,=\,$-0.6$, {\it all} silicon atoms would need to be in the SiC dust
grains. We expect our stars to have somewhat lower metallicity,
corresponding to the peak of the stellar metallicity distribution, say
[Fe/H]\,$\sim -1.0$ and the silicon abundance ([Si/H]) could be $\sim-0.9$. 
Inverting the estimation, the gas-to-dust ratio needs to be $400$ or larger,
to account for the SiC mass fraction of 5\%.  If we consider the uncertainty
in this estimate of the SiC fraction, then the gas-to-dust ratio must be in
the range 800 to 240 to provide a fraction of 10\,\% to 3\% of the dust as
SiC. This may be the first, albeit indirect, indication for a higher
gas-to-dust ratio at significantly sub-solar metallicity. If it is correct,
then the mass-loss rates in AGB stars at these low metallicities could be
even higher than values given in Table\,\ref{table-dusty}. The uncertainty
in the gas-to-dust ratio discussed here largely depends on the uncertainty
of our estimated SiC fraction. Various parameters, but particularly the
optical constants of the dust grains and the dust grain size distribution,
require refinement if further progress is to be made.  Furthermore, the
exact compositions mixed in with the SiC dust grains \citep{Heck07} might
change towards lower metallicity, resulting in different SiC dust masses.
Although it is possible that the gas-to-dust ratio increases at lower
metallicity the actual values remain very uncertain. Due to these
uncertainties, we continue to assume that the gas-to-dust ratio is 200 for
the remaining discussion.

%Although the fraction of SiC from our
%study suggests slightly higher than expected from \citep{Shetrone03}'s
%silicon abundance, this might be partly due to the large uncertainty
%of the our estimated SiC fraction.

%The higher gas-to-dust ratio, if possible, implies that our mass-loss rates derived
%above should be increased by a factor of 2.5 or more. However, it
%also puts the expansion velocity calculated by {\small Dusty} into
%doubt: less (dust) momentum is available per unit of gas mass, by the
%same factor. The two effects are assumed to cancel out.

The dust mass is dominated by amorphous carbon.  The quantity of
amorphous carbon grains is affected by two factors, viz. C/O ratio
\citep{Habing94} and initial metallicity \citep{Zuckerman89,
vanLoon00, Wahlin06}. 
The lack of evidence for lower mass-loss
rates at lower metallicity is explained with a higher C/O ratio towards low
metallicity.  
A strong dependence of mass-loss rate on C/O ratio is reported
from hydrodynamical models for carbon-rich stars \citep{Mattsson07}.  This
is a consequence of the stability of the CO molecule; in carbon stars all
available oxygen atoms are locked into carbon monoxide and excess carbon
atoms are available to form carbon-bearing molecules and amorphous carbon
dust grains. As new carbon atoms reach the surface of these AGB stars, the C/O
ratio increases and the amount of carbon available to form grains also
increases. The C/O abundance of the target
stars in the LMC, the SMC and Fornax have not been
measured directly.  A number of carbon stars in the SMC have 
C/O ratios of 1.05 to 1.1 \citep{deLaverny06}.  That sample probably
contains blue carbon stars, which have little or no circumstellar
envelope.  The C/O ratio increases towards the end of the AGB phase,
but such stars develop a thick circumstellar envelope and hydrostatic
models cannot be used for them, because the molecular lines are
filled in by emission from circumstellar dust grains. 
C/O ratios above 2 are found in LMC and SMC PNe \citep{Leisy96,
Stanghellini05}.  
At metallicities below those of the SMC and LMC, the final
C/O ratios are expected to be even higher due to 
low initial oxygen abundance, and an additional effect of efficient third-dredge up
 \citep{Vassiliadis93, Marigo03, Izzard04}. The
C/O ratios of our targets are therefore likely to significantly exceed 1.1,
leading to the observed efficient amorphous dust production.

This result i.e., the lack of dependence of mass-loss rates on metallicity
for carbon-rich stars, is in contrast to what is found for oxygen-rich stars,
where there is an obvious metallicity dependence \citep{Wood92}.  This is
because dust driven winds in oxygen-rich stars are dominated by the amount
of silicate, and ultimately by the intrinsic silicon elemental abundance
\citep{Bowen91, Willson06}.  The expansion velocities of oxygen-rich shells
are also smaller at low metallicities \citep{Wood92, Zijlstra96,
Marshall04}.  In contrast, carbon stars are strongly affected by carbon
production during the AGB phase; thus the different dependence of mass-loss
rate on metallicity can be understood.

\subsection{Metallicity dependence of molecular bands}
 
The molecular band strengths of carbon-rich stars show a metallicity
dependence \citep{Cohen81, vanLoon99, Matsuura02}.  Recent studies
using infrared spectrometers on the Very Large Telescope
\citep{Matsuura05, vanLoon06} showed that the C$_2$H$_2$ equivalent
widths increase towards lower metallicity, based on samples from the
Galaxy, the LMC and the SMC.  
%In contrast, the 3.55\,$\mu$m HCN
%equivalent widths show a mixed picture; LMC stars show either
%non-detection or HCN equivalent width comparable to the highest
%values found for the Galactic sample.

The carbon-rich stars in the Fornax dSph galaxy follow the same trend of
as those found in similar SMC stars.  For infrared
colours [6.4]$-$[9.3]$>$0.5, the largest C$_2$H$_2$ equivalent widths are
found for stars in the SMC and the Fornax dSph galaxy.  Although Fornax has
a lower metallicity than the SMC, no further metallicity dependence is
seen below the SMC value.

As discussed in the previous section, a higher C/O ratio is likely in
carbon-rich stars at lower metallicity. C$_2$H$_2$ formation relies on excess
carbon atoms after all oxygen atoms are locked into carbon monoxide. 
Thus the higher
C/O ratio results in  the higher abundance of C$_2$H$_2$ at lower metallicity 
\citep{Matsuura05}. 
Fig.\,13 in \citet{Matsuura05} shows that the
fraction of C$_2$H$_2$ in the atmosphere increases drastically up to C/O
$\sim 1.6$, but at ratios higher than this the amount of
C$_2$H$_2$ increases only slowly.  This is because C$_2$H$_2$ is reactive
with carbon resulting in increasingly larger carbon bearing molecules. Thus,
below the SMC metallicity, little metallicity dependence of the C$_2$H$_2$
equivalent width is found, because this molecular abundance becomes
insensitive to the C/O ratio.
%Stars with blue colours
%([6.4]$-$[9.3]$<$0.3--0.5) tend to have artificially large equivalent
%width of C$_2$H$_2$ due to the continuum subtraction problem,
%especially for Fornax dSph stars.

C$_2$H$_2$ is thought to be a parent molecule in the formation of PAHs, as
indicated by chemical models \citep{Allamandola89}. It is still unknown
whether PAHs are formed during the AGB phase or afterwards.  They may
already form in the atmospheres or inner circumstellar envelopes of
carbon-rich AGB stars, where the gas density is reasonably high and chemical
reactions occur easily \citep{Frenklach89}.  The non-detection of PAHs from
AGB stars might be due to a lack of UV radiation, which is probably required
to excite PAHs. PAHs might be photon-processed during post-AGB evolution
\citep{Sloan07}. Alternatively, PAHs may be formed during the post-AGB or
proto-planetary nebula phase, where PAHs are commonly found. The chemistry
in this phase is dominated by strong radiation from the central star
\citep{Woods02, Woods03}. If PAHs are indeed formed during the AGB phase,
an over-abundance of C$_2$H$_2$ in a low metal environment will affect
the growth of these important molecules.

The overabundance of C$_2$H$_2$ at low metallicity will remain through the
post-AGB phase, as is illustrated by the fact that the LMC post-AGB star
SMP\,11 has more abundant carbon-bearing molecules \citep{Bernard-Salas06}
than comparable Galactic post-AGB stars \citep{Cernicharo01}.
\citet{Reyniers07} suggested a lower limit of C/O = 2 for one LMC
post-AGB star.  This overabundance of C$_2$H$_2$ may also occur in an SMC
post-AGB star \citep{Kraemer06}. Thus, at sub-solar metallicity, 
it is likely that C$_2$H$_2$
molecules are abundant in carbon-rich evolved stars.

%Among nearby, star-forming galaxies, PAH emission is suppressed below
%$12+\log ({\rm O/ H}) \sim 8.1$, or approximately
%$log\,z=[Z/\rm{H}]<-0.6$ \citep{Smith07}.  This contrasts with the
%carbon enrichment process of AGB stars. At low metallicity, for
%massive stars the carbon-rich phase (the WR stars) is reduced in
%duration compared to the red supergiant phase \citep{Massey98} and this
%may provide a metallicity-dependent source for the interstellar PAHs,
%however, WR stars are not known to form PAHs. Enhanced UV destruction
%in the dust-poor ISM of low-metallicity galaxies may be important.

\subsection{Gas supply from AGB stars to the ISM of the Fornax dSph galaxy}

From near-infrared colours, \citet{Gullieuszik07} found about 100 AGB
candidates in the centre (18.5$\times$18.5\,arcmin$^{-1}$) of the Fornax
dSph galaxy.  Their infrared colours suggest that at least 20 of these AGB
stars are carbon-rich ($J-K>1.4$ and $K<$14\,mag) \citep[e.g.][]{Cioni03}. 
The reddest carbon stars found by \citet{Gullieuszik07} have 
$J-K=2.4$\,mag.  
An estimate  of mass-loss rate can be obtained from the $J-K$ colors of these 
20 AGB stars of the order of $2\times10^{-7}$\,$M_{\sun}$yr$^{-1}$ or
less, although this number is not very precise owing to the large 
dispersion of the $J-K$ vs mass-loss
rate relation found by \citep{Lebertre97}.
Including our carbon stars, the total mass-loss rate from known
carbon-rich stars in Fornax dSph galaxy is estimated to be about 
$2\times10^{-5}$\,$M_{\sun}$\,yr$^{-1}$ at most, adopting a gas-to-dust ratio
of 200. Three stars in our sample (Fornax 13-23, 12-4, 3-129) contribute
more than two thirds of the total mass-loss rate. Studies of Galactic AGB
stars showed that stars with higher mass-loss rates (above
$1\times10^{-6}$\,$M_{\sun}$\,yr$^{-1}$) dominate the total gas output from
AGB stars \citep{LeBertre01}. Therefore the much bluer AGB stars in Fornax 
are assumed to contribute very little to the total mass lost from the
entire AGB population.

This total mass-loss rate is smaller than the value found for Wolf Lundmark
Melotte (WLM), a dwarf irregular  galaxy  in the Local Group with a similar
metallicity range to Fornax. \citet{Jackson07} estimated the total mass-loss
rate from AGB stars as [0.7--2.4]$\times10^{-3}$\,$M_{\sun}$\,yr$^{-1}$,
which is a factor of 50 larger in WLM than the Fornax dSph galaxy.  
The number ratio of carbon- to oxygen-rich
stars is reported to be 12 in WLM \citep{Battinelli03}; thus the mass-loss
rate from carbon-rich stars is almost identical to that from the entire AGB
population. The visual absolute magnitudes ($M_{V}$) 
of the galaxies are $-14.4$ and $-13.1$ for WLM and Fornax, respectively. 
If $M_{V}$ is a measure of the number of stars in a galaxy,
WLM has a 2 to 3 times larger the number of stars than Fornax has;
the number of stars in the galaxies is
insufficient to explain the difference in total mass-loss rate from
carbon-rich stars.  The assumed gas-to-dust ratio is a factor of 6 to 13
larger in \citet{Jackson07}, but still not sufficient to explain the result.
The difference might be caused by the presence of younger stars in WLM, and
the absence of such stars in Fornax.  In WLM, a recent
increase in star formation rate is indicated starting 1--2.5\,Gyr ago
\citep{Hodge95}.  Young ($<1$\,Gyr) carbon stars dominate the mass output
from AGB stars \citep[e.g.][]{LeBertre01}, and this population is missing in
the Fornax dSph galaxy.

As a dwarf irregular galaxy, WLM has a much larger H{\sc i} mass than the
Fornax dSph: 40\%\ of its total mass compared to $<0.1\% $ for Fornax
\citep{Mateo98}.  A removal of gas from dSph in early phase
is a consequence of this  that difference \citep{Grebel03}.
However, the feedback from the young AGB stars
also accentuates the difference in gaseous contents, on top of the influence
of the galaxy's initial gas content.

%\subsection{Luminosities}

\subsection{Period of Fornax 4-15}

The absolute magnitude of Fornax 4-25 is estimated as
$M_{\rm{bol}}=-4.09$\,mag, using the $K$ magnitude and colour dependent
bolometric correction \citep{Whitelock06}.  The period-luminosity relation
for Galactic carbon stars \citep{Whitelock06} gives $\log~P=2.3$, where $P$
is the period in days, for this luminosity.  The observed period of Fornax 4-15
is $\log~P=2.5$ \citep{Demers87}, thus this Fornax star fits on the
period-luminosity relation for Galactic stars very well. No obvious
metallicity dependence has been found for the AGB period-luminosity
relation.
 This period and mass-loss rate relation for galactic star
 gives mass-loss rate of $10^{-6}$\,$M_{\sun}$\,yr$^{-1}$
 at $\log~P=2.5$. This mass-loss rate is consistent with 
 the upper limit ($< 1.3\times10^{-6}$\,$M_{\sun}$\,yr$^{-1}$)
obtained by this work is consistent.

\section{Conclusions}

 The 5--14\,$\mu$m spectra of AGB stars in the Fornax dSph galaxy confirm
the prevalence of carbon-rich stars at low metallicity
([Fe/H]$\sim-1$).  Stars with  $J-K>3$ show spectra dominated
by C$_2$H$_2$ molecules, together with CO bands and SiC dust excess,
while only a trace of the 7.5\,$\mu$m C$_2$H$_2$ is found in stars with
$J-K<3$. Detection of the SiC dust excess shows that these AGB stars
have developed circumstellar envelopes through mass loss.

We compare our results for Fornax with carbon-rich stars in our own Galaxy,
the LMC and the SMC. The Fornax dSph galaxy is the lowest metallicity galaxy
for which the carbon-rich star population has been studied in the
mid-infrared.  The C$_2$H$_2$ molecular bands, with respect to the infrared
colour, tend to be stronger towards lower metallicity.  No reduction in the
mass-loss rate is found for lower-metallicity carbon stars.  The dust
composition does depend on the metallicity, with a lower mass-fraction of
SiC dust grains, but the amount of amorphous carbon, which constitutes the
dominant fraction of dust grains, seem to be unchanged.  If the fraction of
SiC dust grain is correct, the gas-to-dust ratio could be as high as 400 at
one tenth of the solar metallicity, but a more detailed investigation of gas
mass-loss rates is required. This study of AGB mass loss in a low-metallicity
environment should be broadly useful for other studies of dust and gas
recycling within metal poor galaxies.

Our study reveals that the gas supply from AGB stars into the interstellar
medium is not lower at lower metallicity, but the fraction of SiC dust
grains does appear to be lower. The total amount of gas expelled from AGB
stars is less in the Fornax dSph galaxy than in the dwarf irregular WLM, in
spite of a similar total mass and metallicity. The difference is attributed
to the younger stellar population in WLM.

\section{acknowledgments}
M.M. thanks Prof. Arimoto for discussion about populations in Fornax
dSph galaxy.  M.M. is a JSPS fellow.

\label{lastpage}

\end{document}